\documentclass[letterpaper,english,reprint, aps ,longbibliography]{revtex4-2}
\usepackage[T1]{fontenc}
\usepackage{color}
\usepackage{babel}
\usepackage{float}
\usepackage{amsmath}
\usepackage{amssymb}
\usepackage{graphicx}
\usepackage[unicode=true,pdfusetitle,
 bookmarks=true,bookmarksnumbered=false,bookmarksopen=false,
 breaklinks=false,pdfborder={0 0 1},backref=false,colorlinks=false]
 {hyperref}

\makeatletter

\pdfpageheight\paperheight
\pdfpagewidth\paperwidth

\makeatother

\begin{document}
\title{Self-assembly of Colloids with Competing Interactions Confined in
Spheres}
\author{Ningyi Li}
\thanks{These authors contributed equally.}
\author{Junhong Li}
\thanks{These authors contributed equally.}
\author{Lijingting Qing}
\author{Shicheng Ma}
\author{Yao Li}
\email{liyao@nankai.edu.cn}

\author{Baohui Li}
\affiliation{School of Physics, Key Laboratory of Functional Polymer Materials
of Ministry of Education, Nankai University, and Collaborative Innovation
Center of Chemical Science and Engineering (Tianjin), Tianjin, 300071,
China}
\begin{abstract}
At low temperatures, colloidal particles with short-range attractive
and long-range repulsive interactions can form various periodic microphases
in bulk.In this paper, we investigate the self-assembly behaviour
of colloids with competing interactions under spherical confinement
by conducting molecular dynamics simulations. We find that the cluster,
mixture, cylindrical, perforated lamellar and lamellar structures
can be obtained, but the details of the ordered structures are different
from those in bulk systems. Interestingly, the system tends to form
more perforated structures when confined in smaller spheres. The mechanism
behind this phenomenon is the relationship between the energy of the
ordered structures and the bending of the confinement wall, which
is different from the mechanism in copolymer systems.
\end{abstract}
\maketitle

\section{Introduction}

Self-assembly is a ubiquitous process in which multiple subunits spontaneously
organise into collective and coherent structures or patterns. It plays
an important role in both non-living and living systems, including
polymers, colloids, liquid crystals, biomembranes, and proteins\citep{https://doi.org/10.1002/anie.199611541,Whitesides2002}.
Applications of self-assembly include nanotechnology and nanostructure
producing in polymer systems and colloidal systems\citep{Hu2014,Doerk2017,Miguez2003}. 

Colloids with short-range attractive and long-range repulsive (SALR)
interactions have attracted research interest in recent years. Previous
studies have shown how the attractive and repulsive interactions of
these colloids can be controlled experimentally, by precisely controlling
the paremeters of colloidal suspensions\citep{C8SM00400E,Ruiz-Franco2021}
and very recently dynamic control over the interparticle interactions
by adding magnetic nanoparticles together with external magnetic fields\citep{AlHarraq2022,D3SM00354J}.
The framework of competing interactions also applies to proteins,
amphiphilic and other systems\citep{aa4998b92f23408f9f0f6f307ce5acf0,doi:10.1063/1.2771168,PhysRevLett.94.208301,Sanchez_2005,PhysRevLett.104.165702,PMID:27279005,Ruiz-Franco2021}.
Studies of the self-assembly of colloids with SALR interactions suggest
that short-range interparticle attraction is frustrated by long-range
repulsion, preventing particles from merging and thus leading such
systems to microphase separation and forming ordered phases at low
temperatures, including cluster phase, cylindrical phase, lamellar
phase, and others\citep{Pini2017,Zhuang2016,Serna2019,Zhao_2012}.
These phases are also observed in bulk diblock copolymer systems\citep{yu_cylinder-gyroid-lamella_2005,FD9949800007,PhysRevLett.72.2660,doi:10.1021/ma951138i,doi:10.1021/ma961673y,doi:10.1063/1.1562616,doi:10.1063/1.2140286}.
The remarkable similarity between the bulk phase diagram of colloids
with SALR interactions and that of diblock copolymers has been confirmed
by computer simulations\citep{Zhuang2016} and a Landau-Brazovskii
model has been proposed to explain it\citep{Ciach2013a}.

Self-assembly under confinement often leads to richer patterns and
structures\citep{D0NR04608F,C3SM52821A,Shi2013}. The self-assembly
of diblock copolymers under various confinements has been extensively
studied\citep{Shi2013,Stewart-Sloan2011,yin_simulated_2004,chen_origin_2007,yu_self-assembly_2007,yu_self-assembled_2007,yu_confinement-induced_2006,yu_confinement-induced_2008,chen_microstructures_2008,Yu2007,Zhao2021,Li2009,doi:10.1021/acs.macromol.2c02166,C1SM05947E,doi:10.1021/la202448c,doi:10.1021/ma025559t,10.1063/1.3489685},
among which shows that helical morphologies are observed in systems
confined in cylindrical pores with neutral surfaces\citep{10.1063/1.2178802,10.1063/1.2362818,10.1063/1.3264946,Shi2013}.
Meanwhile, the self-assembly of colloids with SALR interactions confined
in slit pores or cylinders has also been studied\citep{Serna2021b,Serna2021a,Franzini2018},
among which suggests that helical structures are formed in systems
confined in cylindrical pores\citep{Serna2021a}, similar to that
in diblock copolymer systems\citep{10.1063/1.2178802,10.1063/1.2362818,10.1063/1.3264946,Shi2013}.
Structures such as perpendicular lamellae, helices, embedded structures,
perforated lamellar structures and concentric-spherical lamellae are
predicted for diblock copolymers confined inside spherical cavities
in theoretical and simulation studies\citep{Yu2007,chen_microstructures_2008,10.1063/1.3489685,doi:10.1021/ma025559t,doi:10.1021/la200379h,C1SM05947E,doi:10.1021/la202448c,Shi2013}.
SALR fluid on spherical surfaces and SALR colloids forming cylindrical
structures within spherical shells also have been studied recently\citep{D1SM01257F,D3SM00442B}.
However, the self-assembly of colloids with SALR interactions confined
in spherical cavity is not fully understood yet.

In order to elucidate the similarities and differences between the
self-assembly of SALR colloids and diblock copolymers under spherical
confinement, in this paper we investigate the self-assembly of colloids
with SALR interactions confined in spheres via molecular dynamics
simulations. We compare the self-assembly behaviour between our confinement
systems and bulk systems, as well as diblock copolymer systems. The
differences of structures between spherically confined copolymer systems
and bulk copolymer systems are solely due to the surface preference
for a certain block\citep{Shi2013}. In contrast, we reveal that the
differences of structures between our systems and bulk colloidal systems
are due to the bending of the confinement wall.

\section{Model and Methods}

\begin{figure}[H]
\includegraphics[width=1\columnwidth]{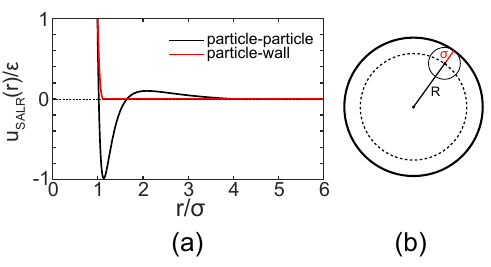}

\caption{Interaction function and illustration for free volume. (a) The LJY
potential (Eq. 1) between a particle and another particle, and the
WCA potential (Eq. 2) between a particle and the wall employed. (b)
Illustration for free volume. The radius of the free volumn of the
particle centres is $R=R_{w}-\sigma$, where $R_{w}$ is the radius
of the spherical wall.\label{fig:function}}
\end{figure}

The interaction between colloidal particles is described by an effective
SALR potential, which is the addition of a Lennard-Jones (LJ) potential
plus a screened electrostatic interaction represented by the Yukawa
potential (LJY):

{\footnotesize{}
\begin{equation}
u_{\mathrm{SALR}}\left(r_{ij}\right)=4\epsilon\left[\left(\frac{\sigma}{r_{ij}}\right)^{12}-\left(\frac{\sigma}{r_{ij}}\right)^{6}\right]+\frac{A}{\left(r_{ij}/\lambda\right)}\exp\left(-r_{ij}/\lambda\right)
\end{equation}
}where $r_{ij}$ is the distance between particles $i$ and $j$;
$\epsilon$ and $\sigma$ are the usual LJ parameters. The parameter
$A$ measures the strength of the electrostatic interaction, and $\lambda$
is the Debye screening length. The parameters are given the same values
as in the bulk systems\citep{Serna2019}. The potential function for
$\epsilon=1.6$, $\sigma=1.0$, $A=0.65$ and $\lambda=2.0$ is shown
as the black curve in Fig. \ref{fig:function}(a). In particular,
we truncate the potential at $r_{\mathrm{cut}}/\sigma=4.0$ to improve
computational efficiency. In the following, we use $\sigma$ and $\epsilon$
as the units of distance and energy respectively.

The confinement is imposed by placing a spherical wall with a radius
of $R_{w}$. Particles interact with the wall via truncated Lennard-Jones
potential (WCA)\citep{Andersen1971}:

{\footnotesize{}
\begin{equation}
u_{iw}\left(r_{iw}\right)=\begin{cases}
4\epsilon_{w}\left[\left(\frac{\sigma_{w}}{r_{iw}}\right)^{12}-\left(\frac{\sigma_{w}}{r_{iw}}\right)^{6}\right]+\epsilon_{w} & r_{iw}<2^{1/6}\sigma_{w}\\
0 & r_{iw}\geq2^{1/6}\sigma_{w}
\end{cases}
\end{equation}
}where $\epsilon_{w}/\epsilon=1.0$, $\sigma_{w}/\sigma=1.0$, and
$r_{iw}$ is the distance between particle $i$ and the spherical
wall $R_{w}$. The WCA potential between the particles and the spherical
wall is shown as the red curve in Fig. \ref{fig:function}(a). The
total energy of this system is given by

\begin{equation}
u_{\mathrm{tot}}=\sum_{i=1}^{N-1}\sum_{j>i}^{N}u_{\mathrm{SALR}}\left(r_{ij}\right)+\sum_{i=1}^{N}u_{iw}\left(r_{iw}\right)
\end{equation}

We employ molecular dynamics (MD) by using the open-source package
LAMMPS\citep{Plimpton1995,THOMPSON2022108171}. We choose a confinement
radius range of $R/\sigma=4$\textendash $12$ and a system density
of $0.8$ or less for our simulations. We employ the simulated annealing
method to accelerate the equilibration of the systems and obtain the
structures at low temperature\citep{Kirkpatrick1983}. The temperature
is controlled by a Langevin thermostat with a relaxation time of $100\mathrm{d}t$,
where the time step is $\mathrm{d}t=0.005\sqrt{m\sigma^{2}/\epsilon}$.
The simulations are run for $10^{7}$ time steps for equilibration
at each temperature. The annealing process starts from the temperature
$T^{*}=k_{B}T/\epsilon=1.0$ to $T^{*}=0.4$ with a temperature step
of $\mathrm{d}T^{*}=0.1$, and then continues from $T^{*}=0.4$ to
$T^{*}=0.1$ with a temperature step of $\mathrm{d}T^{*}=0.003$.

The radius of the free volume for particles $R$ is illustrated in
Fig. \ref{fig:function}(b). The number density is calculated based
on free volume, which is defined as $\rho^{*}=\frac{N\sigma^{3}}{4\pi R^{3}/3}$.
The radial density distribution $\rho{}_{r}^{*}\left(r\right)$ is
defined by

\begin{equation}
\rho{}_{r}^{*}(r)=\frac{N\left(r,r+\Delta r\right)}{\frac{4}{3}\pi\left(r+\Delta r\right)^{3}-\frac{4}{3}\pi r^{3}}
\end{equation}
where $N\left(r,r+\Delta r\right)$ is the number of particles between
$r$ and $r+\Delta r$. We set $\Delta r=0.1\sigma$ in our calculation.
All plots of radial density distribution are averaged from 10000 configurations,
each of which is sampled every $100\mathrm{d}t$.

The various snapshots of the systems are plotted by the open-source
visualisation software OVITO\citep{Stukowski2010}. In order to clearly
visualise the global structures, the alpha-shape method\citep{article}
and the Gaussian density method\citep{10.2312:PE:EuroVisShort:EuroVisShort2012:067-071}
are employed, which are surface mesh functions in OVITO.

The bond-orientation order parameters $q_{l}$ and the hexatic order
parameter $\Phi_{6}$ for particles are calculated\citep{PhysRevB.28.784,PhysRevB.19.2457}.
For each particle $i$, bond-orientation order parameter $q_{l}$
is a real number defined as $q_{l}=\sqrt{\frac{4\pi}{2l+1}\sum_{m=-l}^{m=l}\bar{Y}_{lm}\bar{Y}_{lm}^{*}}$,
where $\bar{Y}_{lm}=\frac{1}{N_{\mathrm{nn}}}\sum_{j=1}^{N_{\mathrm{nn}}}Y_{lm}\left(\theta\left(\boldsymbol{r}_{ij}\right),\phi\left(\boldsymbol{r}_{ij}\right)\right)$.
Here $Y_{lm}\left(\theta,\phi\right)$ are the spherical harmonics,
bond vector $\boldsymbol{r}_{ij}$ is from particle $i$ to one of
its nearest neighbours, particle $j$, and $N_{\mathrm{nn}}$ is the
number of nearest neighbours. We choose $l=6$ and $l=8$ in our calculation.
The data of $q_{6}$ and $q_{8}$ are obtained by the LAMMPS \textit{compute}
command. 

For each particle $i$ in a 2D system, hexatic order parameter $\Phi_{6}$
is defined as $\Phi_{6}=\frac{1}{N_{\mathrm{nn}}}\sum_{j=1}^{N_{\mathrm{nn}}}\cos\left[6\theta\left(\boldsymbol{r}_{ij}\right)\right]$,
where $N_{\mathrm{nn}}$ is the number of nearest neighbours, bond
vector $\boldsymbol{r}_{ij}$ is from particle $i$ to one of its
nearest neighbours, particle $j$, and the angle $\theta$ is formed
by the bond vector $\boldsymbol{r}_{ij}$ and the given $x$ axis.
Though the systems we investigate are 3D, we calculate the $\Phi_{6}$
of a particular particle by projecting the particle and its nearest
neighbours in a spherical layer onto its tangent plane. 

\section{Results and Discussions}

\begin{figure*}
\includegraphics[width=0.8\paperwidth]{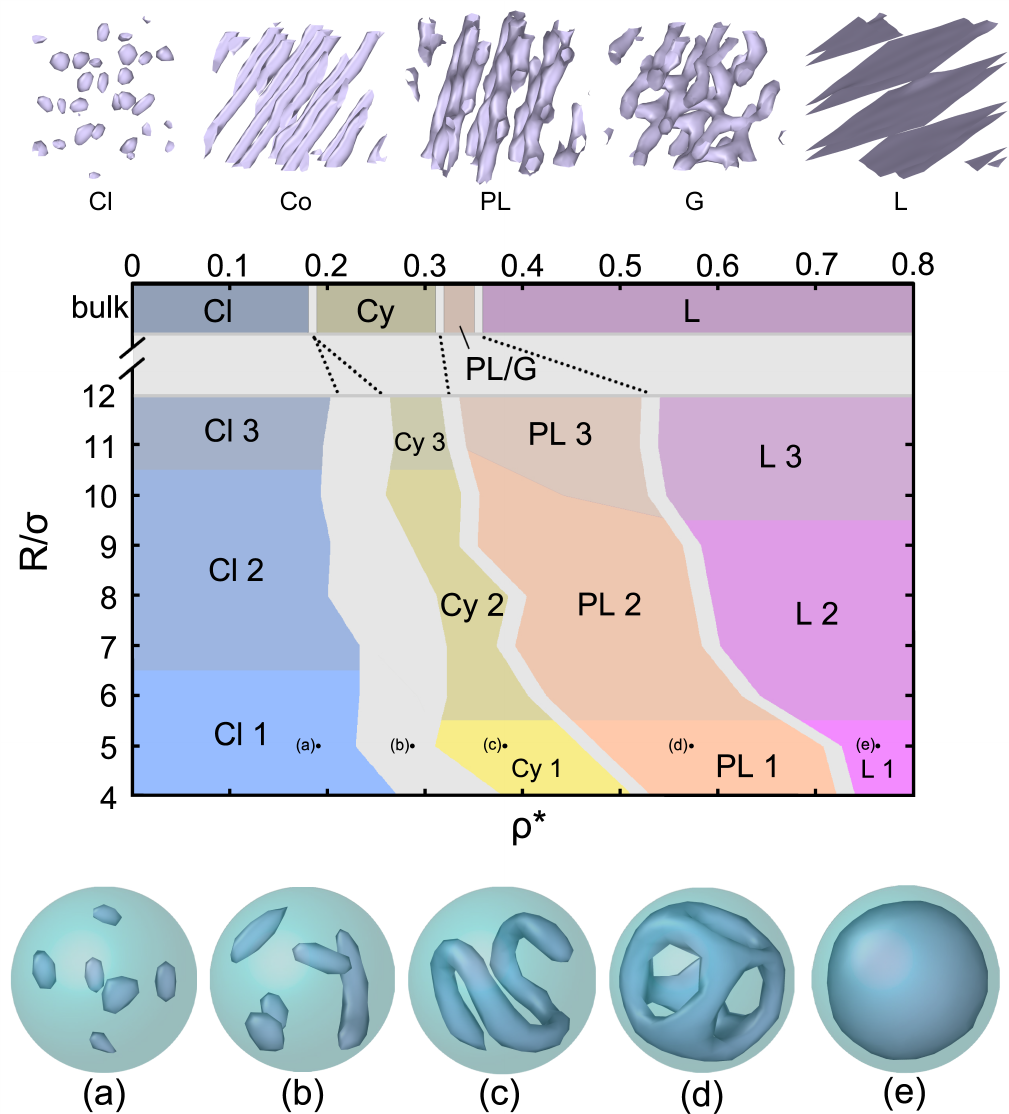}

\caption{Phase diagram and typical structures. Top row: Surfaces of each typical
structure in bulk systems obtained by the alpha-shape method. The
perforated lamellar (PL) structure and Gyroid (G) structure can occur
under the same conditions when the density is appropriate. Mid panel:
Phase diagram with density and confinement radius as coordinates at
$T^{*}=0.1$, where different colours indicate different structures.
The results of the bulk system are also shown. The phase behaviour
shows four structures in order of increasing density: cluster (Cl),
cylindrical (Cy), perforated lamellar (PL) and lamellar (L). The number
of layers distributed radially increases with larger confinement radius,
which is shown by the numbers in the diagram. The phase boundary lines
are drawn in gray colour which represent the co-existence zone of
the phases. All the phase boundaries roughly reflect the trend that
the smaller the confinement radius is, the more to the right it is.
Bottom row: Surfaces of each typical structure in the confinement
system obtained by the alpha-shape method. Representive snapshots
are taken with the parameters indicated by the solid dots in the phase
diagram.\label{fig:phaseDiagram}}
\end{figure*}

The phase diagram is shown in Fig. \ref{fig:phaseDiagram}. The density
range we investigate is from $0$ to $0.8$, with a density separation
$\Delta\rho^{*}<0.02$ between two adjacent simulated parameters.
The systems spotaneously form ordered structures at low temperature
for chosen parameters. 

As density increases, we obtain four ordered structures of cluster,
cylindrical, perforated lamellar and lamellar in turn. The structure
formed at each thermodynamic state is identified from surface snapshots.
Spherelike clusters are identified as cluster strcutures. Long cylinders
or rings with no or only one small cluster in the centre are identified
as cylindrical structures. Concentric structures with three-fold or
more than three-fold junctions are identified as perforated lamellar
structures. We identify the structures as lamellar when there are
no holes observed on the surface.

For the points near the phase boundary on the phase diagram, we performed
10 simulations, for the other points on the phase diagram, we performed
6 simulations. Points where the same structure consistently appears
are identified as corresponding to that structure (i.e. the coloured
regions in the phase diagram Fig \ref{fig:phaseDiagram}), while points
where different structures emerge are identified as co-existence zones
(i.e. the gray regions in the phase diagram Fig \ref{fig:phaseDiagram}).
We select the midpoint between two adjacent points representing different
structures as the phase boundary on the phase diagram. 

The phase boundaries generally tend to be at higher density with smaller
confinement radius. When we compare our systems with the bulk systems,
we find that there is no wide co-existence zone from cluster to cylinder
in the bulk systems. But there is a wide co-existence region between
cluster structures and cylindrical structures in the phase diagram,
where the mixture of clusters and short cylinders are observed. In
bulk systems, the region of the perforated lamellar structures in
the phase diagram is very narrow. The gyroid structures are observed
in bulk systems. However, the gyroid structures are not observed in
either colloidal systems or polymer systems under confinement.

\begin{figure}
\includegraphics[width=1\columnwidth]{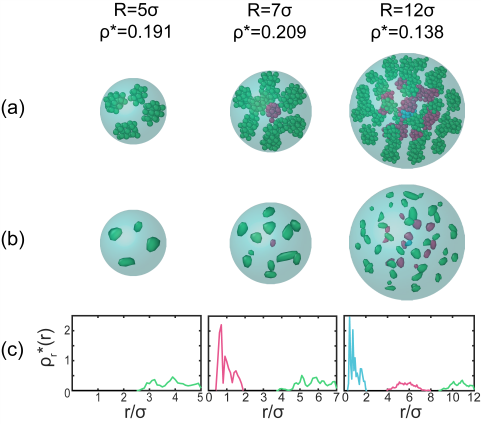}\caption{Typical configurations and radial density distribution of the cluster
structure. Snapshots are shown in (a) for the location of the particles
and in (b) for the surfaces obtained by the alpha-shape method, with
cluster structures formed under different parameters. (c) Radial density
distribution for the corresponding cluster configurations, averaged
from 10000 configurations, each of which is sampled every $100\mathrm{d}t$.
Snapshots and plots of radial density distribution are colour-coded
according to the radially distributed layer. \label{fig. cluster-layer}}
\end{figure}

\textbf{Cluster.} We show typical snapshots of cluster structures
in Fig. \ref{fig. cluster-layer}(a) and (b). As the confinement radius
of the system increases, the radial distribution of clusters form
one-, two- and three-layer structures, which can be clearly recognized
by the radial density distribution in Fig. \ref{fig. cluster-layer}(c).

\begin{figure}
\includegraphics[width=1\columnwidth]{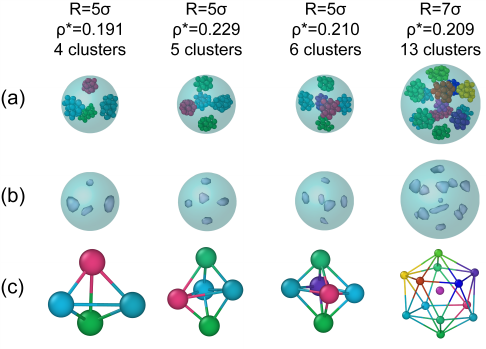}\caption{Typical configurations of clusters arranged as the vertices of Platonic
solids (or convex regular polyhedrons). From left to right is tetrahdron,
double tetrahdron, octahedron, icosahedron, respectively. Snapshots
are shown in (a) for the locations of the particles and in (b) for
the surfaces obtained by the alpha-shape method. Each sphere in (c)
corresponds to the center of a cluster shown in (a). \label{fig:clusterStructure}}
\end{figure}

We also find that the outermost clusters are arranged into vertices
of regular Platonic solids (convex regular polyhedrons) at low temperatures
in some cases (Fig. \ref{fig:clusterStructure}). This is very similar
to diblock copolymer systems, and the packing of geometric solids
under spherical confinement\citep{doi:10.1021/acs.macromol.2c02166,doi:10.1073/pnas.1524875113}.
Among these polyhedrons, some are identified as ortho-tetrahedrons,
ortho-octahedrons, ortho-icosahedrons, and others, as shown in Fig.
\ref{fig:clusterStructure}. While in bulk systems, the cluster phase
forms a lattice structure close to face-centred cubic (FCC) or body-centred
cubic (BCC)\citep{Serna2019}.

\begin{figure}
\includegraphics[width=1\columnwidth]{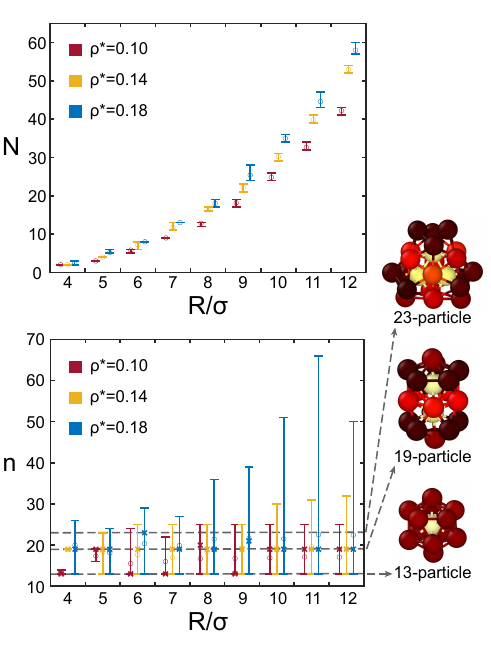}\caption{Statistical fluctuation about the number and size of clusters formed.
The top panel shows the minimum and maximum of the number of clusters
formed, and the bottom panel shows the minimum and maximum of cluster
size, in 3 different densities. The average value is shown by circles,
and the mode for cluster size is shown by crosses. Three dashed lines
in the bottom panel represent the mostly observed cluster sizes. Clusters
with 13, 19, and 23 particles are mostly observed in our system, of
which the shapes are shown at the right.\label{fig:cluster_size_=000023_shape}}

\end{figure}

We count the number and size of clusters from multiple simulations
at each thermodynamic parameters. For all of our simulation results,
no clusters with less than 13 particles are observed. When the density
is $\rho^{*}=0.18$, some big clusters emerge for it is close to the
co-existence zone between cluster and cylinder. Though there is some
big clusters with more than 30 particles, they are always very few.
We notice that clusters with 13, 19, and 23 particles are usually
the most in our system. The shapes of clusters with 13 particles,
19 particles and 23 particles are regular, as shown in Fig. \ref{fig:cluster_size_=000023_shape},
which is the same as the results in the bulk system\citep{Serna2019}.

\begin{figure}
\includegraphics[width=1\columnwidth]{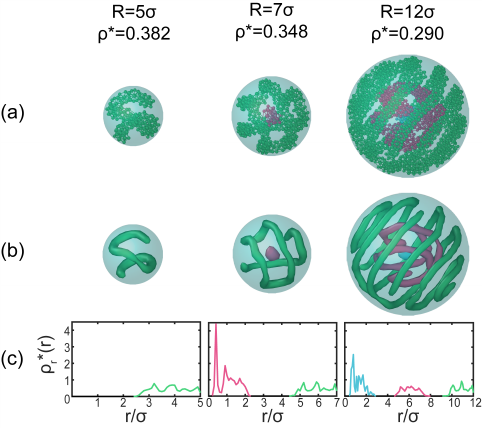}\caption{Typical configurations and radial density distribution of the cylindrical
structure. Snapshots are shown in (a) for the location of the particles
with bonds added and in (b) for the surfaces obtained by the alpha-shape
method, with cylindrical structures formed under different parameters.
Bonds are added between the nearest neighbour particles in (a) for
facilitating observation. (c) Radial density distribution for the
corresponding configurations, averaged from 10000 configurations,
each of which is sampled every $100\mathrm{d}t$. Snapshots and plots
of radial density distribution are colour-coded according to the radially
distributed layer (same as in Fig. \ref{fig. cluster-layer}).\label{fig: Column1}}
\end{figure}

\textbf{Cylindrical. }We show typical snapshots of cylindrical structures
in Fig. \ref{fig: Column1}(a) and (b). Clearly, the radial distribution
of cylinders form one-, two- and three-layer structures as the confinement
radius increases, which can also be recognized by the radial density
distribution in Fig. \ref{fig: Column1}(c). The cylinders are curved
along the spherical wall, which is different from bulk systems where
the cylinders mostly maintain a straight shape\citep{Serna2019}.

\begin{figure}
\includegraphics[width=1\columnwidth]{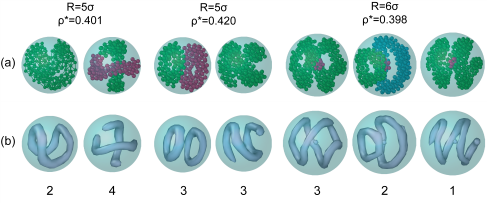}\caption{Configurations of ring, saddle-shape and bent cylindrical structure.
Snapshots are shown in (a) for the location of particles with bonds
added and in (b) for the surfaces obtained by the alpha-shape method.
There are three groups of snapshots, and snapshots in each group are
taken under same parameters with different random seeds in the simulation.
The 1st and the 5th are saddle-shape structure. The 2nd, the 4th and
the 7th are bent column structure. The 3rd and the 6th are ring structure.
Each number at the bottom row shows the occurrence frequency of a
specific structure among 6 simulations. \label{fig:columnSaddle}}
\end{figure}

Multiple cylindrical structures can be obtained at the same thermodynamic
state, which is similar to confined diblock copolymer systems\citep{doi:10.1021/acs.macromol.2c02166}.
Bent cylindrical strutures, ring structures and saddle-shape structures
can be formed, with snapshots shown in Fig. \ref{fig:columnSaddle}.
We perform 6 simulations with different random seeds for each thermodynamic
state. The average energy per particle is nearly equal for different
cylindrical structures obtained at the same thermodynamic state. Examples
include that the average energy per particle is around $\left(-1.48\pm0.03\right)\epsilon$
when $R=5\sigma$ and $\rho^{*}=0.401$, around $\left(-1.45\pm0.03\right)\epsilon$
when $R=5\sigma$ and $\rho^{*}=0.420$, and around $\left(-1.41\pm0.03\right)\epsilon$
when $R=6\sigma$ and $\rho^{*}=0.398$. Different cylindrical structures
are obtained at different frequencies in different thermodynamic states.
Configurations and occurrence frequencies of frequent structures at
some specific thermodynamic states are shown in Fig. \ref{fig:columnSaddle}.
The morphology of the saddle-shape structure is similar to the semiflexible
ring polymers under spherical confinement\citep{Guven2012a}. As the
density increases, the saddle-shape ring might bend further, or forming
multiple rings, or becoming longer cylinders (Fig. \ref{fig:columnSaddle}). 

\begin{figure}
\includegraphics[width=1\columnwidth]{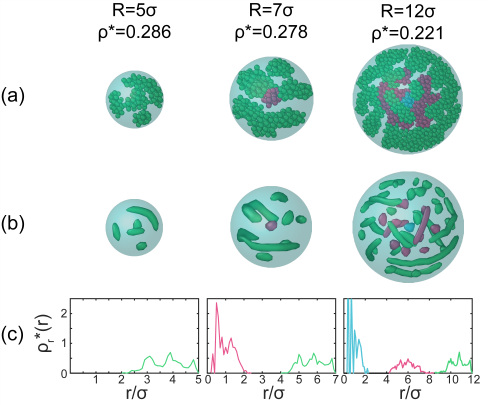}\caption{Typical configurations and radial distribution of the mixture structure
of clusters and cylinders. Snapshots are shown in (a) for the location
of the particles and in (b) for the surfaces obtained by the alpha-shape
method, with mixture structures formed under different parameters.
(c) Radial density distribution for the corresponding configurations
(truncated at $\rho_{r}^{*}\left(r\right)=2.5$), averaged from 10000
configurations, each of which is sampled every $100\mathrm{d}t$.
Snapshots and plots of radial density distribution are colour-coded
according to the radially distributed layer (same as in previous figures).\label{fig: Mixture}}
\end{figure}

When the density is between the density of forming the cluster and
cylindrical structures, mixture structures of clusters and cylinders
are formed, with typical snapshots shown in Fig. \ref{fig: Mixture}(a)
and (b). It is a very wide co-existence region of clusters and cylinders.
The radial distribution of mixtures forms one-, two- and three-layer
structures as the confinement radius increases, which is the same
as previously discussed in cluster structures and cylindrical structures.
The radial density distribution is plotted in Fig. \ref{fig: Mixture}(c).

\begin{figure}
\includegraphics[width=1\columnwidth]{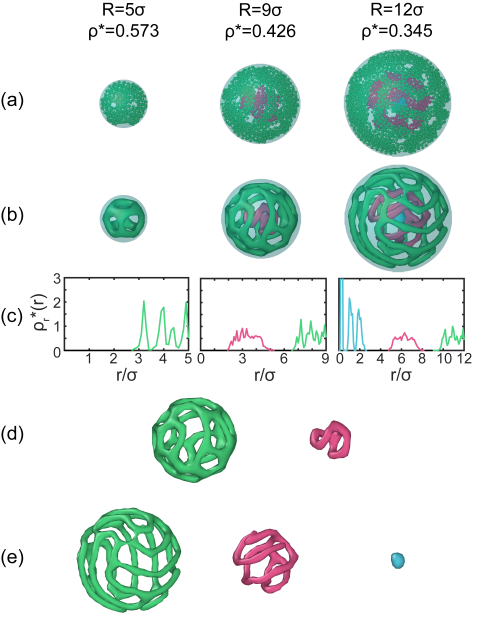}\caption{Typical configurations and radial density distribution of the perforated
lamellar structure. Snapshots are shown in (a) for the location of
the particles with bonds added and in (b) for the surfaces obtained
by the alpha-shape method. (c) Radial density distribution for the
corresponding configurations (truncated at $\rho_{r}^{*}\left(r\right)=3$),
averaged from 10000 configurations, each of which is sampled every
$100\mathrm{d}t$. (d) Seperated snapshots of surfaces when $R=9\sigma$
and $\rho^{*}=0.426$. Each layer shows a similar perforated lamellar
structure. (e) Seperated snapshots of surfaces when $R=12\sigma$
and $\rho^{*}=0.345$. The innermost particles form a cluster. Snapshots
and plots of radial density distribution are colour-coded according
to the radially distributed layer (same as in previous figures).\label{fig:(a):perforated layer fig}}
\end{figure}

\textbf{Perforated Lamellar.} With density larger than that of forming
the cylindrical structures, perforated lamellar structures are formed.
Typical snapshots are shown in Fig. \ref{fig:(a):perforated layer fig}(a)
and (b). As the density increases, the cylinders are connected to
each other. These structures are similar to the structures of diblock
copolymers confined in spheres\citep{Yu2007,Li2009}. Clearly, systems
exhibit multi-layer structures as the wall radius gradually increases,
which can also be recognized by the radial density distribution shown
in Fig. \ref{fig:(a):perforated layer fig}(c). We note that at $R=9\sigma$
and $\rho^{*}=0.426$, the system is a two-layer structure (Fig. \ref{fig:(a):perforated layer fig}(d)).
Both the outer layer and the inner layer are perforated lamellar structure.
While at $R=12\sigma$ and $\rho^{*}=0.345$, the innermost particles
form a cluster (Fig. \ref{fig:(a):perforated layer fig}(e)).

\begin{figure}
\includegraphics[width=1\columnwidth]{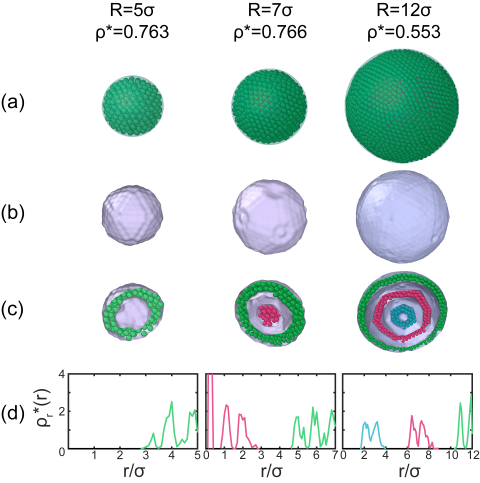}\caption{Typical configurations and radial density distribution of the lamellar
structure. Snapshots are shown in (a) for the location of the particles
and in (b) for the surfaces obtained by the Gaussian density method.
(c) Sliced snapshots for the corresponding configurations in (b).
(d) Radial density distribution for the corresponding configurations
(truncated at $\rho_{r}^{*}\left(r\right)=4$), averaged from 10000
configurations, each of which is sampled every $100\mathrm{d}t$.
Snapshots and plots of radial density distribution are colour-coded
according to the radially distributed layer (same as in previous figures).\label{fig:layer}}
\end{figure}

\begin{figure}
\includegraphics[width=1\columnwidth]{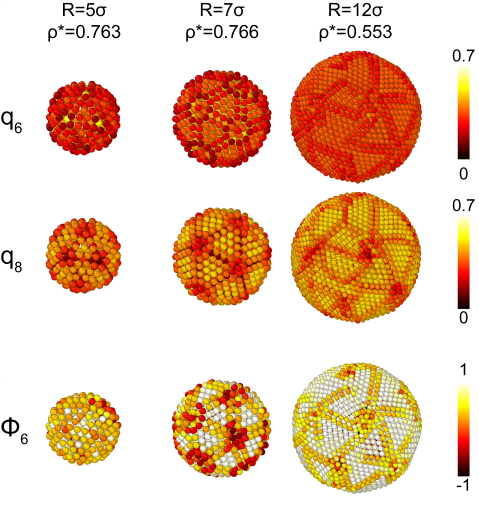}

\caption{Snapshots of the lamellar structure colour-coded according to the
order parameters. The top and mid rows are according to $q_{6}$ and
$q_{8}$ respectively, with the colour bar range from $0$ to $0.7$.
The bottom row is according to $\Phi_{6}$ order, with the colour
bar range from $-1$ to $1$. The high order parameters indicate most
parts of the layers are crystallised except those unavoidable defects.
\label{fig:LayerOrder}}
\end{figure}

\begin{figure}
\includegraphics[width=1\columnwidth]{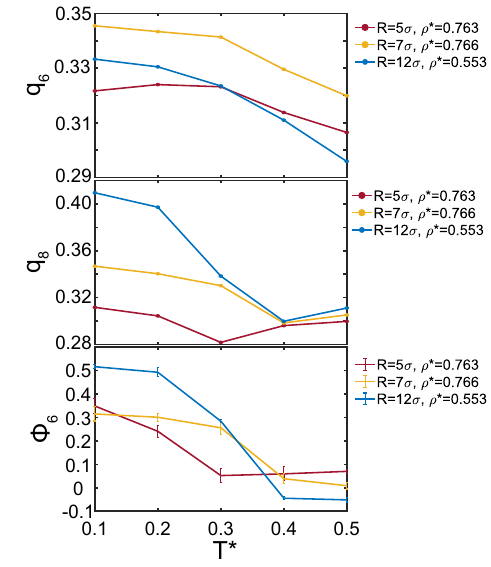}\caption{Three structure order parameters as functions of temperature $T^{*}$,
for 3 different confined radii and densities in Figs. \ref{fig:layer}
and \ref{fig:LayerOrder}. The top shows $q_{6}$ order, the mid shows
$q_{8}$ order, and the bottom shows $\Phi_{6}$ order. The error
bars (minimum and maximum) of $\Phi_{6}$ are shown in the plot. The
sizes of error bars (minimum and maximum) of $q_{6}$ and $q_{8}$
are smaller than the size of the data dots.\label{fig:order_temperature}}
\end{figure}

\textbf{Lamellar.} When the systems are denser, lamellar structures
are formed. The lamellae crystallise at low temperatures (as shown
in Fig. \ref{fig:layer}(a) and (b), Fig \ref{fig:LayerOrder}), which
is not observed in confined diblock copolymer systems\citep{Shi2013}.
According to the plots shown in Fig. \ref{fig:order_temperature},
when the temperature is over $T^{*}=0.4$, the order parameter $\Phi_{6}$
is nearly $0$ and $q_{6}$ and $q_{8}$ are also small, which suggests
that the system is disordered at high temperatures. The order parameters
$\Phi_{6}$, $q_{6}$ and $q_{8}$ significantly increase at low temperatures,
meaning that systems become ordered and the lamellae crystallise except
for unavoidable defects. The local shape of the spherical lamella
is nearly flat to form an energetically favourable structure (Fig.
\ref{fig:layer}(b)). Similar to experimental and simulated results
involving other colloids or hard geometric solids, crystallised lamellae
form ortho-icosahedrons\citep{Wang2018,Wang2019,Wang2021,Chen2021,WangDa2018}.
We can find that there are obvious patterns of defects, satisfying
the topologic requirement.

In bulk systems, the lamellar structure behaves as flat lamellae parallel
to each other\citep{Serna2019}. In our systems, the particles form
a concentric spherical lamellar structure, which is also observed
in spherically confined diblock copolymer systems\citep{Yu2007,Shi2013}.
As the wall radius increases, the system gradually exhibits a multilayer
structure (Fig. \ref{fig:layer}(c)). Each layer will form a same
lamellar structure when the density and the confinement radius is
appropriate. The thickness of each lamella is 2 or 3 particles, as
shown in the snapshots in Fig. \ref{fig:layer}(c) and the radial
density distribution in Fig. \ref{fig:layer}(d). The thickness of
each spherical lamella is not fixed. It is influenced by the size
and density.

\begin{figure}
\includegraphics[width=1\columnwidth]{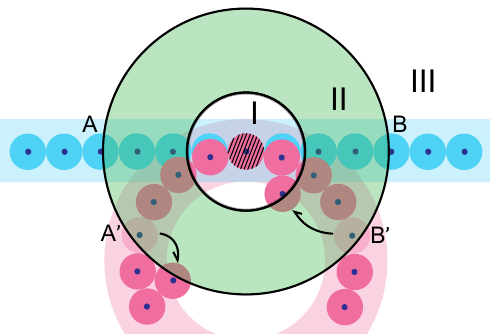}\caption{Illustration for the mechanism of bending. This figure is a sliced
illustration for 3D system. The thickness of each lamella is 2 or
3 particles, which is illustrated by the light blue and light red
area. Roughly, the blue circles represent particles of a straight
lamella in bulk systems, while the red circles represent particles
of a bending lamella in spherical confinement systems. Particles A
and A', B and B' represent the same particle in bulk and under spherical
confinement respectively. The green-coloured region II represents
the spherical-shell-shape space of positive potential energy, while
the part inside (region I) represents the space of negitive potential
energy, given by the shaded particle at the centre. The shaded particle
do not interact with outside particles (region III). As the flat lamella
bends along with the spherical wall, some particles enter the region
II from region III (e.g. particles A' and B'), increasing the average
energy per particle. These particles will be repelled to region III
like particle A', or be attracted to region I like particle B', for
reducing the additional potential energy, which makes the lamella
perforated.\label{fig:demonstration-1}}
\end{figure}

\begin{figure}
\includegraphics[width=1\columnwidth]{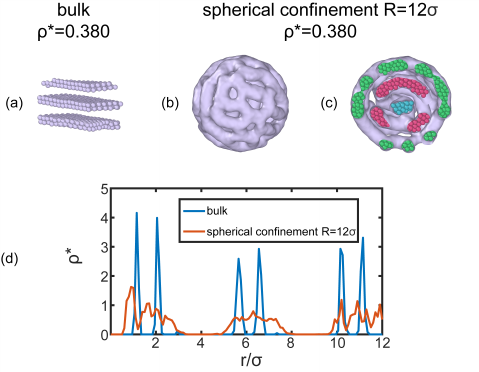}

\caption{Comparison between the bulk system and the spherical confinement system
with the same density at $\rho^{*}=0.380$. (a) Snapshot of the bulk
system showing the location of the particles, with a lamellar structure
formed. (b) Surfaces of the spherical confinement system with $R=12\sigma$
obtained by the Gaussian density method, with a perforated lamellar
structure formed. (c) Sliced snapshot of (b). (d) Density distribution
of the bulk system (along the direction perpendicular to the lamellae),
and radial density distribution of the system under confinement, averaged
from 1000 configurations, each of which is sampled every $100\mathrm{d}t$.\label{fig:discussion}}
\end{figure}

The phase boundaries generally tend to be at greater density with
smaller wall radius (Fig. \ref{fig:phaseDiagram}). We proposed the
mechanism for explaining this phenomenon, with an illustration shown
in Fig. \ref{fig:demonstration-1}. The layer composed of particles
is flat in bulk systems, while it will bend when it is confined in
a spherical wall. As the layer bends along with the spherical wall,
some particles in the region III in Fig. \ref{fig:demonstration-1}
enter the region II, increasing the average energy per particle. Particles
entering region II will be attracted by the shaded particle into the
lower-energy region I, or be repelled to the lower-energy truncated
region III, therefore the layer will be thicker and more perforated,
reducing the additional positive potential energy. Consequently, more
perforated lamellar structures are formed at the density where lamellar
structures are formed in bulk systems. Concretely, when the density
is $\rho^{*}=0.380$, a lamellar structure is formed in the bulk system,
while a perforated lamellar structure is formed when it was confined
in a sphere with $R=12\sigma$ (Fig. \ref{fig:discussion}(a), (b),
and (c)). Due to the same reason, colloids confined in spheres tend
to form more cylindrical structures at the density where perforated
lamellar structures are formed in bulk systems.

Furthermore, the smaller the wall radius is, the greater the curvature
of the lamella becomes, which means that more particles enter the
region II, resulting more increase in average energy per particle.
Thus, the system is energetically more favourable for forming a perforated
structure when the confinement radius is smaller, which results in
the phase boundaries tilting to the right.

Our simulations show that colloidal systems under spherical confinement
tend to form more perforated lamellar structures than colloidal systems
in bulk (Fig. \ref{fig:discussion}(a), (b), and (c)). Though our
systems form perforated lamellar structures with a smaller outer surface,
the perforated lamella is thicker than the lamellae in bulk (which
can be clearly recognised by the density distribution shown in Fig.
\ref{fig:discussion}(d)), suggesting that the number of particles
in one layer being almost unchanged. Also, the structure of each layer
is almost the same. It suggests that, compared with the bulk system,
when layer in a confinement system become bent, the particles at the
layer relocate within the same layer as shown in the Fig. \ref{fig:demonstration-1}
and form more perforated and thicker structures, rather than relocating
to another layer.

In terms of lamellae in bulk systems, the distance between lamellae
is between $3\sigma$ and $4\sigma$ (see the blue curve in Fig. \ref{fig:discussion}(d)).
It means that diferent particles at different lamellae interact with
each other in the end of the repulsive region with very small potential
energy, or in the cut-off region (see the potential function in Fig.
\ref{fig:function}(a)). If more perforated and thicker lamellar structures
are formed in the bulk system, the distance between different lamellae
become closer, around $2.5\sigma$, and more particles will interact
in the repulsive region with greater potential energy. These particles
get much greater positive potential energy from interlayer interaction,
thus it is not energetically favourable to form perforated structures
in bulk. However, as previously discussed, forming perforated and
thicker structures can reduce the additional intralayer potential
energy in confinement systems. The reduction of intralayer energy
is greater than the increase of interlayer energy brought by becoming
thicker. This competition between interlayer energy and intralayer
energy results in more perforation under confinement but less perforation
in bulk. 

The confinement wall we use is neutral which only has volume exclusion.
Colloids confined in spheres form more perforated structures than
in bulk, while the diblock copolymers confined in neutral spherical
walls are not obviously perforated than in bulk\citep{Shi2013}. The
reason for this difference is that, there is not a mechanism of bending
in confined diblock copolymer systems. In confined diblock copolymer
systems, the interaction between the wall and the block dominates,
and blocks tend to relocate themselves among different layers. The
self-assembly structures are significantly different according to
the wall surface preference for the block\citep{Yu2007,Shi2013}.

\begin{figure}
\includegraphics[width=1\columnwidth]{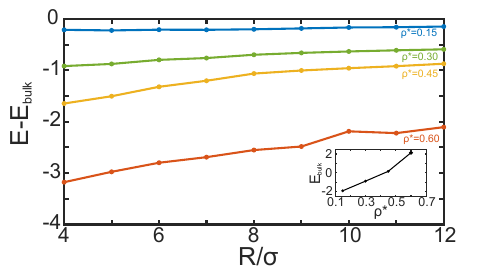}\caption{Difference of the average energy per particle between the bulk system
and the spherical confinement system. The solid blue, green, yellow
and orange curves are the average energy per particle in confinement
systems of different confinement radii minus that in the bulk system,
for densities $\rho^{*}=$ 0.15, 0.30, 0.45, and 0.60 respectively.
The inserted figure is a plot of energy versus density in the bulk
system. \textcolor{black}{The }fluctuation of the energy is smaller
than the solid dots in the plot.\label{fig:Energy}}
\end{figure}

\begin{figure}
\includegraphics[width=1\columnwidth]{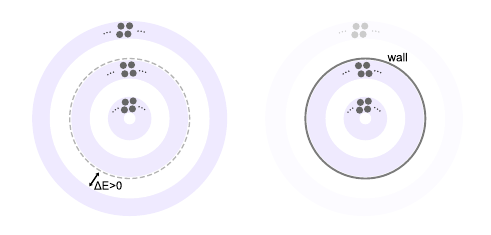}\caption{Illustration for the decrease in energy brought by confinement. This
is merely an illustration, and the dots in the figure do not refer
to any specific structures. Each coloured ring represents a lamella
with a thickness of 2 or 3 particles. Left: Illustration of the bulk
system. There is a positive potential energy $\Delta E$ between the
outer particles and the more outer particles. As the radius goes to
infinity, the spherical layer becomes a flat layer. The case of a
bulk system will be equivalent to the case of this illustration. Right:
Illustration of the confinement system. With the addition of the spherical
confinement wall, there are no more outer particles in the system
and therefore no positive potential energy $\Delta E$. \label{fig:demonstration}}
\end{figure}

We measure the average energy per particle for dentities $\rho^{*}=0.15$,
$0.30$, $0.45$ and $0.60$, at $T^{*}=0.1$ for different confinement
radii. At the same density, the average energy per particle in our
systems is below that in bulk systems, which is shown in Fig. \ref{fig:Energy}.
The explanation for this pheomenon is illustrated in Fig. \ref{fig:demonstration}.
In both bulk and confinement systems, the distance between lamellae
(or cylinders, or clusters) remains around $3\sigma$ (see e.g. Fig.
\ref{fig:discussion}(d)). Particles at this distance repel each other
with a positive potential energy. As a result, there is a positive
potential energy between the outer particles and the inner particles.
In spherical confinement systems, the energy given by outside of the
outermost particles vanished (Fig. \ref{fig:demonstration} right
column), resulting in a decrease in average energy per particle compared
with the bulk systems (Fig. \ref{fig:demonstration} left column).
At the same time, the smaller the confinement radius is, the larger
the ratio of the surface particle number to the number of all particles
becomes, meaning that more energy per particle vanished. This causes
the energy diagram to reflect a trend that the smaller the wall radius
is, the smaller the average energy per particle of the system is.

The confinement causes additional positive energy generated by the
bending of the layer. But the vanished energy is significantly greater
than the additional energy brought by bending. Therefore, the overall
average energy per particle for our confinement systems is smaller
than that for the bulk systems. 

\section{Conclusion}

In this article we have studied colloidal systems with competing interactions
confined in spheres. Our results showed that, in spherical confinement
systems of particles with SALR interactions, the cluster, cylindrical,
perforated lamellar, and lamellar structures are formed at low temperature.
The configuration of self-assembled clusters in some cases fits the
vertices of Platonic solids. More structures such as ring structures
are formed under confinements compared with the bulk systems\citep{Serna2019}.
We also found that the average energy per particle is below that in
bulk systems. Crystallisation occurs on the surface at appropriate
densities, which is not observed in diblock copolymer systems\citep{Yu2007,Shi2013}.

Moreover, we found that the phase boundaries tend to be at higher
density with smaller confinement radius. Our systems tend to form
perforated structures when confined in smaller spheres, which is different
from diblock copolymer systems\citep{Shi2013}. Perforated lamellar
structures are formed in spherical confinement systems at densities
where lamellar structures are formed in the bulk systems. Layers composed
of particles bend along the wall in confined SALR colloidal systems,
which increases the average energy per particle. Perforated lamellar
structures are energetically favourable under spherical confinement.
Due to the same reason, column structures are formed at densities
where perforated lamellar structures are formed in the bulk systems.
However, in confined diblock copolymer systems, the differences of
structures compared with the bulk systems are almost only due to the
interaction between the wall and the block\citep{Shi2013}. The intralayer
relocation of the colloidal particles makes the layer becomes perforated
and thicker, while the possible changes of polymers only caused by
interlayer relocation. The mechanism of structural changes in colloidal
systems under spherical confinement is completely different from that
in copolymer systems. The studies on SALR fluid on spherical surfaces
and SALR colloids in spherical shells focused on quasi-two-dimensional
systems, characterized by a single layer of particles\citep{D1SM01257F,D3SM00442B}.
Thus, it is impossible to investigate the pheononon about multiple
layers, for example the competition between interlayer energy and
intralayer energy results more perforated lamellar structure formation
in the current work.

We hope this study helps to the design of particle self-assembly under
confinement especially the utility of the curvature of the confinement
wall, with promising applications in nanostructure designs and understanding
living systems, such as porous structure formations.

\section*{Author Contribusions}

Conceptualisation, Y. L.; methodology, N. L and J. L.; validation,
N. L. and J. L.; formal analysis, N. L., J. L., and Y. L.; investigation,
N. L., J. L., and Y. L.; data curation, N. L. and J. L.; writing\textemdash original
draft, N. L., J. L.; writing\textemdash review and editing, N. L.,
J. L., L. Q., S. M., Y. L., B. L.; visualisation, N. L. and J. L.;
supervision, B. L. and Y. L.; project administration, B. L. and Y.
L.; funding acquisition, B. L. and Y. L..

\section*{Conflicts of Interest}

There are no conflicts to declare.
\begin{acknowledgments}
This work was supported by the National Natural Science Foundation
of China (12275137), Fundamental Research Funds for the Central Universities,
Nankai University (63221053, 63231190). We thank Ho-Kei Chan and Peng
Tan for helpful discussions, and Jun Zhong for helping on data visualisation.
\end{acknowledgments}

%\begin{thebibliography}

%\end{thebibliography}

%\bibliographystyle{rsc}
%\bibliography{SALR}
%apsrev4-2.bst 2019-01-14 (MD) hand-edited version of apsrev4-1.bst
%Control: key (0)
%Control: author (8) initials jnrlst
%Control: editor formatted (1) identically to author
%Control: production of article title (0) allowed
%Control: page (0) single
%Control: year (1) truncated
%Control: production of eprint (0) enabled
%

\end{document}